\documentclass[letterpaper, 10 pt, conference]{ieeeconf}  % Comment this line out if you need a4paper

\IEEEoverridecommandlockouts                              % This command is only needed if 
\usepackage{graphicx}
\usepackage{amsmath}
\usepackage{cite}
\usepackage{tabulary}
\usepackage{amssymb}
\usepackage{subcaption}
\usepackage{multirow}
\usepackage{booktabs}
\usepackage{hyperref}

\title{\LARGE \bf
Identifying trac\'e alternant activity in neonatal EEG using an inter-burst detection approach}

\author{Sumit A. Raurale$^{1,2}$ \textit{Member, IEEE}, Geraldine B. Boylan$^{1,2}$, Gordon Lightbody$^{1,3}$ and \\ John M. O'Toole$^{1,2}$ \textit{Member, IEEE} % <-this % stops a
  \thanks{This work is supported by the Wellcome Trust (209325/Z/17/Z).  
    JMOT is supported by Science Foundation Ireland (15/SIRG/3580 and
    18/TIDA/6166).}% <-this
  \thanks{$^{1}$INFANT Research Centre, Ireland. 
    {\tt\small (sumit.raurale@ucc.ie)}}%
  \thanks{$^{2}$Department of Paediatrics \& Child Health, University College Cork,
    Ireland.}%
  \thanks{$^{3}$Department of Electrical \& Electronic Engineering, University College
    Cork, Ireland.}%
}

\begin{document}

\maketitle
\thispagestyle{empty}
\pagestyle{empty}

%%%%%%%%%%%%%%%%%%%%%%%%%%%%%%%%%%%%%%%%%%%%%%%%%%%%%%%%%%%%%%%%%%%%%%%%%%%%%%%%
\begin{abstract}
Electroencephalography (EEG) is an important clinical tool for reviewing sleep–wake cycling in neonates in intensive care. Trac\'e alternant (TA)---a characteristic pattern of EEG activity during quiet sleep in term neonates---is defined by alternating periods of short-duration, high-voltage activity (bursts) separated by lower-voltage activity (inter-bursts). This study presents a novel approach for detecting TA activity by first detecting the inter-bursts and then processing the temporal map of the bursts and inter-bursts. EEG recordings from 72 healthy term neonates were used to develop and evaluate performance of 1) an inter-burst detection method which is then used for 2) detection of TA activity. First, multiple amplitude and spectral features were combined using a support vector machine (SVM) to classify bursts from inter-bursts within TA activity, resulting in a median area under the operating characteristic curve (AUC) of 0.95 (95\% confidence interval, CI: 0.93 to 0.98). Second, post-processing of the continuous SVM output, the confidence score, was used to produce a TA envelope.
This envelope was used to detect TA activity within the continuous EEG with a median AUC of 0.84 (95\% CI: 0.80 to 0.88). 
%This envelope was able to detect the TA activity within a long-duration EEG with a Cohen's kappa of 0.57, F1-score of 79.7\%, overall accuracy of 81.9\% and a median AUC of 0.84 (95 CI: 0.80 to 0.88). 
These results validate how an inter-burst detection approach combined with post processing can be used to classify TA activity. 
Detecting the presence or absence of TA will help quantify disruption of the clinically important sleep--wake cycle. 
\end{abstract}

%Identification of trac\'e alternant is an important stage for the classification of term EEG in neurologically compromised neonates. 

\section{Introduction}
\label{introduction}
Sleeping is the primary activity of newborns. Disturbances to the sleep--wake cycle can
provide valuable insights into neurological development and maturation
\cite{Korotchikova1}.
Electroencephalography (EEG) provides a continuous measurement of electrical brain
activity that is well suited to analysing sleep--wake cycling in the neonate
\cite{Korotchikova1}. 
But around-the-clock EEG monitoring and review in most neonatal intensive care units (NICU) is not practical. Automated EEG analysis could help by presenting the physician
with useful and timely clinical information about brain function.  

%In addition, the neonatal EEG in the premature infant is in constant development. Especially the duality between the two sleep states evolves during development. Below 31 weeks postmenstrual age (31w PMA) the different sleep stages can not be distinguished, while at 32 to 36 weeks there is a maximal separation between the 2 sleep states. After 37 weeks, different sleep stages start flourishing in the EEG signal with the first seeds of Tracé Alternant, which is fully present at full-term age. This study
%The sleep period includes active sleep (AS) similar to wakefulness, as both show continuous traces and quiet sleep (QS) exhibits discontinuous traces, consisting of burst cycles of high amplitude separated by inter-burst intervals \cite{Korotchikova1}. 

Sleep--wake cycling is evident in healthy term neonates within hours of birth
\cite{Korotchikova1}. This cycle can be classified into 4 behavioural states: active
sleep, quiet sleep, indeterminate sleep, and wakefulness. Quiet sleep itself consists of 2 different EEG patterns, high-voltage slow wave ($<$4 Hz) activity and trac\'e alternant (TA) activity. TA activity is characterised by high voltage waveforms, typically 50 to 150 $\mu$V peak-to-peak, separated by lower-voltage waveforms, typically between 25 to 50 $\mu$V peak-to-peak \cite{ACNS}. We refer to the higher voltage activity as bursts and the lower voltage activity as inter-bursts. Each waveform lasts from a few seconds up to approximately 10 seconds \cite{ACNS}.
% Although we specify voltage and time ranges for these waveforms,
% they are never absolute and may vary for different neonates and even evolve over
% time. From past experience of reviewing hundreds or possibly thousands of neonatal EEGs,
% an expert clinical physiologist can incorporate experience with the guidelines to identify these waveforms. 

%Trac\'e discontinu activity, present in preterm EEG, over time evolves into the TA in term EEG.  

%The AS and QS are further sub-divided as (1) mixed frequency active sleep, (2) low voltage irregular active sleep, (3) high voltage slow wave quiet sleep, and (4) trac\'e alternant (TA) quiet sleep which shows similar discontinuous trace as in QS with equal length of bursts and inter-bursts patterns in full-term neonates EEG \cite{Dereymaeker1}. The trac\'e alternant is an important physiological marker of normal function and maturation \cite{Dereymaeker1,Korotchikova1}.

%Thus, automated detection of this distinctive EEG patterns will be helpful in clinical purposes. 

There has been a limited amount of work on automating the detection of different sleep states in neonatal EEG. Dereymaeker \emph{et al.} \cite{Dereymaeker2} presented a method to detect quiet sleep and active sleep for preterm infants. Both Pillay \emph{et al.} \cite{Pillay} and Ansari \emph{et al.} \cite{Ansari} developed different methods to detect components of quiet and active sleep states, including TA activity, using a combined cohort of preterm and term infants with EEG recorded at term-equivalent age.  
Turnbull \emph{et al.} \cite{Turnbull} focused solely on detecting TA activity in term EEG. This study explored the potential for using the discrete wavelet transform, using a small dataset of 20 EEG segments from 6 neonates.

As TA activity is an essential component of quiet sleep, it is therefore an important neurophysiological marker of normal function and maturation \cite{Dereymaeker1,Korotchikova1}. We aim to develop a method that detects the presence or absence of TA activity within an EEG recording. TA activity, with its quasi-periodic sequence of bursts and inter-bursts, contrasts sharply with other EEG activity. To develop the TA detector, we first construct a detector to differentiate between bursts and inter-bursts. Next, we process the output of the burst--inter-burst detector and use this processed envelope to distinguish between TA activity and non-TA activity in the continuous EEG. We intend to use this TA module as part of an algorithm that grades neonatal EEG for hypoxic ischemic encephalopathy described in \cite{embc}. A TA activity detector, combined with a grading algorithm \cite{embc}, could help distinguish normal from abnormal EEG.

\begin{figure*}[!h]
	\centering
	%\vspace{0.15cm}
	\includegraphics[width=0.94\linewidth]{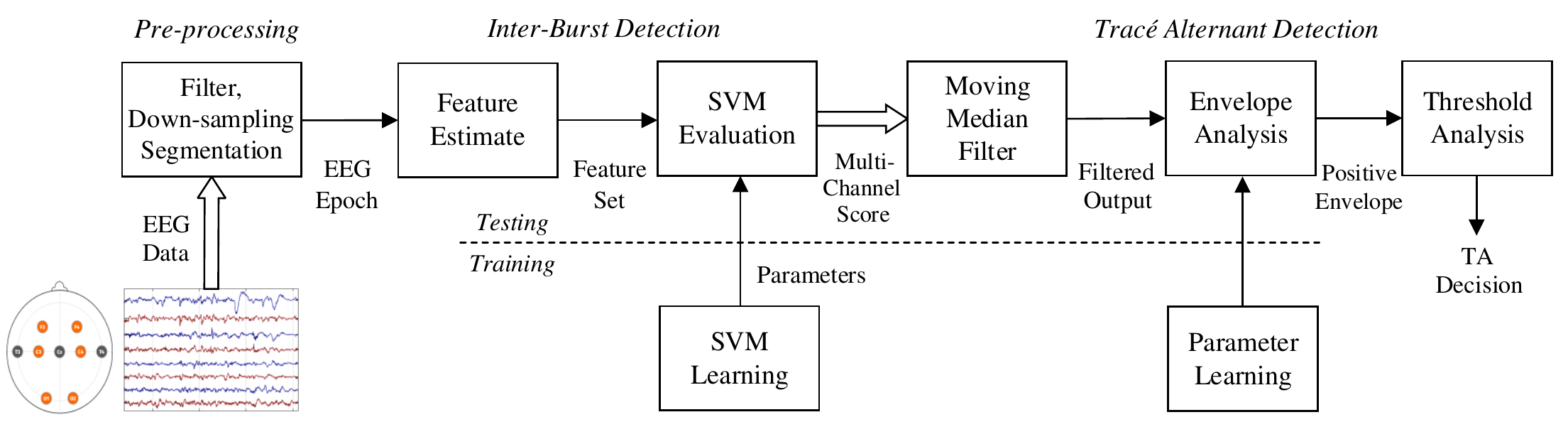}
	%\vspace{0.15cm}
	\caption{Overview of proposed system for detecting trac\'e alternant activity. SVM: support vector machine; TA: trac\'e alternant.}
	\label{system}
\end{figure*}

\section{Methods}

We first develop an automated method to segment TA activity into bursts and inter-bursts. We do this by extracting multiple features and combine using a machine learning method. Second, we process the classifier's decision function when applied to the whole EEG, which includes both TA and non-TA activity, to generate a score to detect TA. The overall structure of the proposed system is shown in Fig. \ref{system}.

\subsection{EEG Data and Pre-processing}

EEG was recorded from term infants using a NicoletOne EEG system in Cork University Maternity Hospital, Cork, Ireland. Informed and written parental consent was obtained before EEG recording. The study was approved by the Clinical Ethics Committee of the Cork Teaching Hospitals. EEG recordings started as soon as possible after birth and continued for up to 1--2 hours to include different sleep states. Five scalp electrodes were used over the frontal and temporal regions with reference in the midline (Cz). We analysed the EEG using a 4-channel bipolar montage, derived from these electrodes as F3-T3, F4-T4, T4-Cz and Cz-T3.

\begin{figure}[!h]
	\centering
	\includegraphics[width=1.0\linewidth]{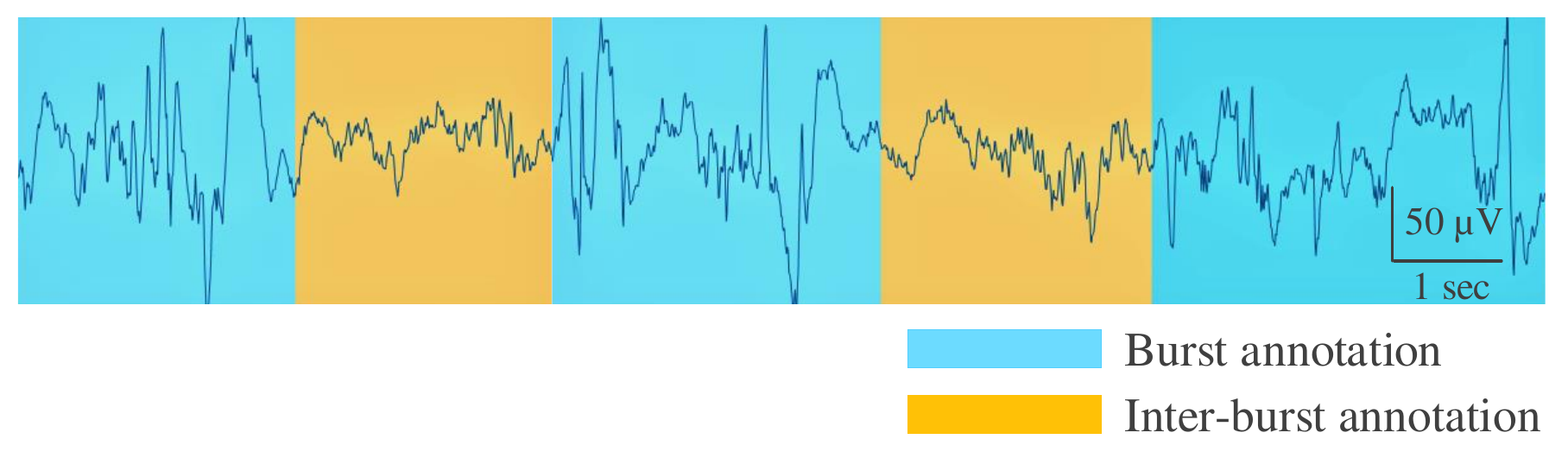}
	\caption{Annotations of bursts and inter-bursts within trac\'e alternant activity for 1 channel of EEG.}
	\label{fig_eegex}
\end{figure}

A subset of 72 EEGs, from a total of 91 EEGs, were selected for this study based on presence of $>$2 min of continuous TA activity. This dataset is detailed in Korotchikova \emph{et al.} \cite{Korotchikova1}.
An EEG expert reviewed and annotated instances of TA activity \cite{Korotchikova1}. Within the TA activity all bursts and inter-bursts were also annotated. Fig. \ref{fig_eegex} shows an example of annotated bursts and inter-bursts for 1 channel.

EEG data were sampled at 256 Hz during recording and electrode impedance was maintained below 5 k$\Omega$. Movement artefacts, defined as the absolute value of EEG $>$1,500 $\mu$V, were removed. EEG was low-pass filtered to 30 Hz using an finite-impulse response (FIR) filter of length 4,001 samples and then down-sampled to 64 Hz.

\subsection{Burst and Inter-burst Detection}

As part of the proposed system, we first developed a method to classify bursts and inter-bursts present within TA activity. Because TA activity is a maturation of the trac\'e discontinue activity present in preterm EEG, we started by modifying a feature set developed to detect bursts and inter-bursts in preterm EEG \cite{Toole2}. These features capture differences in amplitude and spectral shape across four modified frequency bands 0.5–4 Hz, 4–7 Hz, 7–13 Hz, and 13–30 Hz, and one frequency-weighted energy measure called the envelope--derivative operator within the 0.5--10 Hz band \cite{Toole3}. EEG signals $x(n)$ are bandpass filtered into the $i$-th frequency band using a 5th-order Butterworth filter, resulting in $x_i(n)$. The filters are implemented using a forward–backwards procedure to obtain a zero-phase response \cite{Toole2}. The features are defined, for finite signal $x_i(n)$ of length $N$, as follows.

\vspace{0.15cm}
\noindent 1. Envelope: quantifies power in the signal by computing the median of the signal envelope $e_i(n)$. The envelope is defined as
\begin{equation}
e_i(n)  =  \vert z_i(n) \vert ^2 = \vert x_{i}(n)+j H [x_{i}(n)]\vert ^2
\label{equ1}
\end{equation}
where $z_i(n)$ is the analytic associate of $x_i(n)$, $H[\cdot]$ represents the Hilbert transform, and $j$ represents the imaginary unit. 

\vspace{0.15cm}
\noindent 2. Fractal dimension: time-domain approach to quantify the complexity of the signal. Here we use the Higuchi method, which first estimates curve length at the $k$th-scale as,
\begin{equation}
\begin{aligned}
L_{m}(k) = {} & \dfrac{(N-1)}{\lfloor (N-m)/k \rfloor k^2} \;\; \times \\ 
& \sum_{i=1}^{\lfloor (N-m)/k \rfloor} \vert x[m+ik] - x[m+(i-1)k] \vert 
\end{aligned}
\label{equ4}
\end{equation}
over $m=1,2,...,k$ using the entire frequency range 0.5–30 Hz. Curve length $L(k)$, at scale $k$, is then computed as the mean value of $L_{m}(k)$ over all $m$ values. This process is iterated for different scale values $k$. The slope of a line fit to the points (log $k$, log $L(k)$) provides an estimate of fractal dimension.

\vspace{0.15cm}
\noindent 3. Relative spectral power: quantifies spectral shape by assessing the relative power of the $i$-th frequency band,
\begin{equation}
P_i = \dfrac{\sum_{k\in i} \vert X(k) \vert ^2}{P_{\text{total}}}
\label{equ2}
\end{equation}
where $X(k)$ is the discrete Fourier transform (DFT) of $x(n)$, $P_{\text{total}}$ is the total spectral power over the 0.5–30 Hz range, and notation $\sum_{k\in i}$ represents summation over the $i$-th frequency band. 

\vspace{0.15cm} 
\noindent 4. Measure of spectral fit: also quantifies spectral shape. The line $\widehat{Y}(l)=c_1 + c_2 l$ approximates the log--log spectrum $Y(l)$ and measure of fit $r^2$, defined as
\begin{equation}
r_{i}^2 = 1-\dfrac{\sum_{l\in i} \left[ Y(l)-\widehat{Y}(l) \right]^2 }{\sum_{l\in i} \left[ Y(l)-\dfrac{1}{N_i} \sum_{l\in i} Y(l)\right]^2 }
\label{equ3}
\end{equation}
is used as the feature. $Y(l)$ is the log of $|X(k)|^2$ at log frequency $l$. Each $i$-th frequency band is fitted separately. 

\vspace{0.10cm}
\noindent 5. Instantaneous frequency: yet another feature that quantifies spectral shape. The feature is computed as the median of the instantaneous frequency $f_i(n)$, where $f_i(n)$ is estimated using the central-finite difference as,
\begin{equation}
f_i (n)=\dfrac{f_s}{4\pi}\left\lbrace \left[\phi _i (n+1)-\phi _i (n-1)\right] ~\text{mod}~2\pi \right\rbrace 
\label{equ6}
\end{equation}
with phase function $\phi _i (n)=\text{arg}[z_{i}(n)]$, where $z_{i}(n)$ is the analytic signal from (\ref{equ1}) for the $i$-th frequency band.

\vspace{0.10cm}
\noindent 6. Envelope–derivative operator: quantifies the frequency-weighted energy of the signal. This non-negative measure combines both amplitude and frequency as \cite{Toole3},
\begin{multline}
\Gamma(n)=\dfrac{1}{4}[x^2 (n+1)+x^2 (n-1)+h^2 (n+1)+h^2 (n-1)] + \\ [x(n+1)x(n-1)+h(n+1)h(n-1)]
\label{equ5}
\end{multline}
where $h(n) = H[x(n)]$ is the Hilbert transform of $x(n)$. 

Features 1 to 5 are generated on short-duration epochs with 75\% overlap. Time-domain features (1 and 2) use 1-second epochs; spectral features (3, 4, and 5) use 2-second epochs. The envelope--derivative operator (feature 6) is computed on the whole signal. Features 1, 3, 4, and 5 are computed for the 4 different bands, resulting in a total of 18 features. Feature selection criteria of an area-under the operating characteristic curve (AUC) $>$0.6 was implemented for each feature individually within a leave-one-subject-out (LOSO) cross-validation. 

This feature set was combined using a linear support vector machine (SVM), in keeping with the preterm burst-detection method on which this feature set is based \cite{Toole2}. In addition, 2 other models were tested: a Fisher discriminate analysis model and a random forest model. The random forest hyper-parameters---number of trees and maximum number of levels in each decision tree---were selected within a nested cross-validation scheme. We also implemented a radial basis function SVM but found little difference in performance during initial testing compared to the linear SVM and therefore we only consider the linear SVM here.

Computer code (Matlab) for the inter-burst detector trained on all EEG records is available at
\url{https://github.com/sumitraurale/interburst\_detector}.

\subsection{TA Detection}

\begin{figure}[!h]
	\centering
	\includegraphics[width=0.98\linewidth]{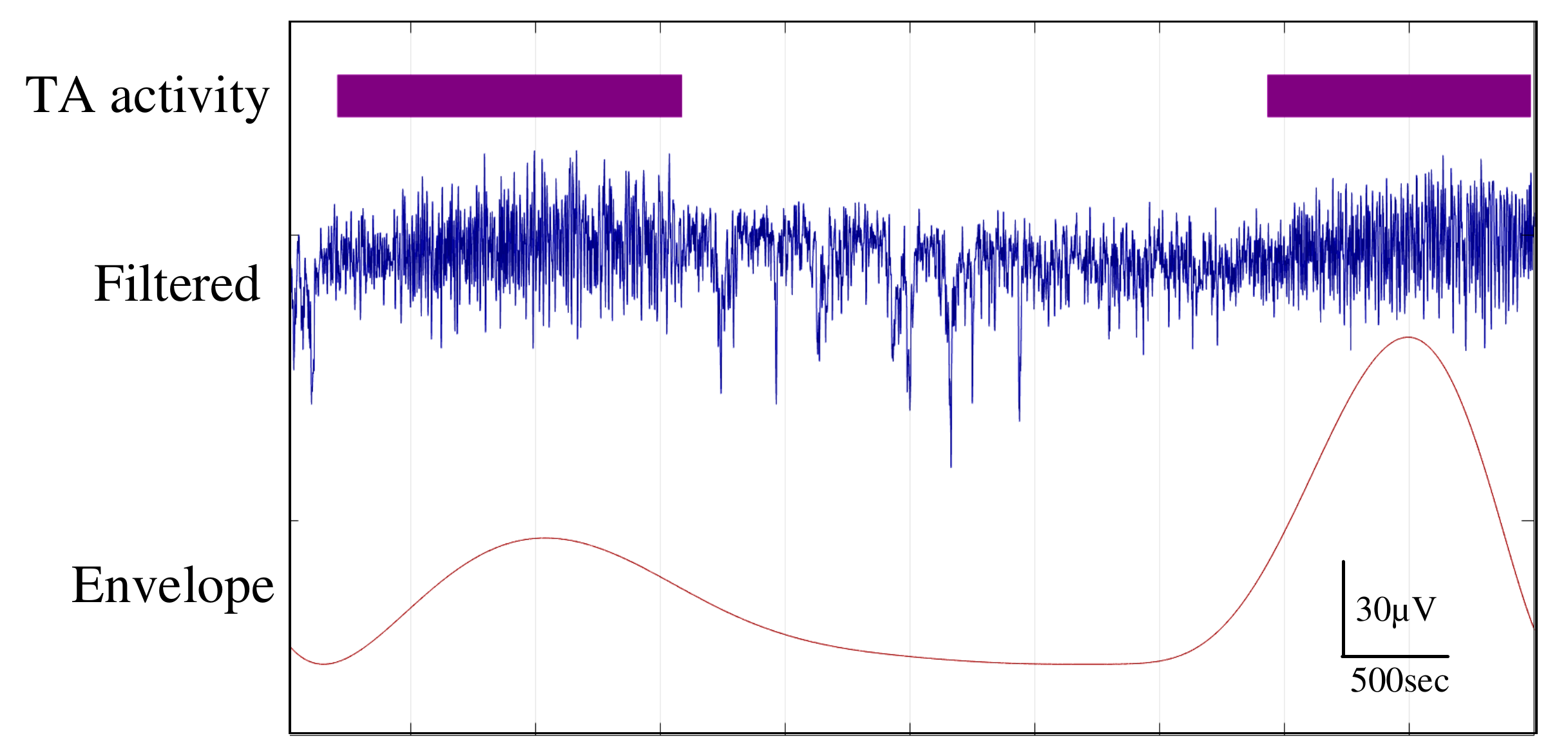}
	\caption{TA envelope for 1.3 hours of EEG. Filtered confidence score from the inter-burst detector and smoothed envelope function depicted in relation to TA activity.}
	\label{fig_ta}
\end{figure}

We use the continuous-valued output of the burst detector, the confidence score, to differentiate TA activity from other EEG activity. The process is as follows. First, we apply a low-pass smoothing process to the score to suppress the higher-frequency noise and outliers. For this we use a 3-second moving-median filter. We produce a summarised filtered score by averaging across channels. Second, we then apply an envelope estimation method to the filtered confidence score. Local maxima are computed on the score with a parameter that specifies the minimum separation between consecutive peaks. These peaks are then joined using spline interpolation to create a smooth envelope function. This envelope is in turn a confidence score of the presence or absence of TA activity. The minimum separation parameter for the local maxima is optimised during training over the range (2.5, 50) seconds with step size of 2.5-seconds. Each EEG is segmented into 20 minute epochs and only those epochs with either full TA activity or full non-TA activity are considered. Fig. \ref{fig_ta} shows an example of the filtered confidence score from the burst detector and the envelope, highlighting the smoothing effect of the envelope estimation process. 

For training and testing the inter-burst and TA detector, we use LOSO cross-validation. Feature selection for the inter-burst detector and parameter optimisation for the TA envelope method is generated within the LOSO cross-validation. 

%Based on 72 infants considered in study, system performance was observed by training SVM classifier within the burst detector with 71 infants EEGs and testing on the one left out. This was progressed until all 72 infant's EEGs with only burst and inter-burst data were tested to determine the performance of the burst (and inter-burst) detector. The trained SVM model is then tested on the complete LOSO EEG recording to evaluate trac\'e alternant performance.

\section{Results}
\label{expresult}
\begin{figure}[!h]
	\centering
	\includegraphics[width=0.85\linewidth]{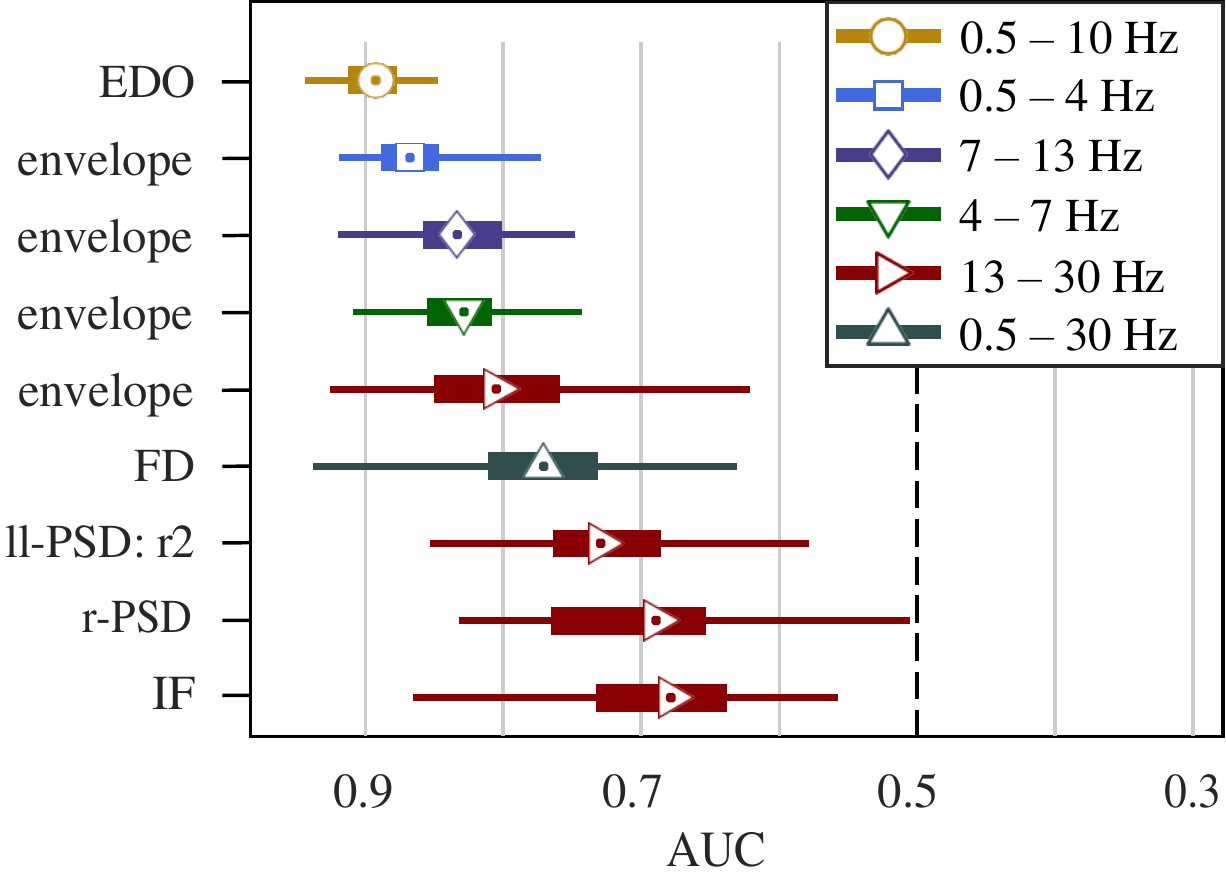}
	\caption{Detection performance for individual features of the inter-burst detector, calculated in different frequency bands. Dots represent median values, thick lines inter-quartile range, and thin lines 95th percentile range of AUC values for the 72 infants. EDO: envelope--derivative operator; r-PSD: relative power spectral density (PSD); ll-PSD: r2: linear fit ($r^2$) to log--log PSD; IF: instantaneous frequency; FD: fractal dimension. }
	\label{fig_featauc}
\end{figure}

The same feature set of 9 features were selected at each of 72 LOSO iterations. Individual feature performance for these features are illustrated in Fig.~\ref{fig_featauc} based on an AUC for each EEG record.

AUC values, comparing classifiers, for the inter-burst detection component of the proposed system are illustrated in Table \ref{auc_vclass}. Although there is little difference in performance the linear SVM outperforms both the FDA and random forest classifiers. Thus we proceeded with this SVM to generate the TA envelope and detector.

\begin{table}[!h]
\renewcommand{\arraystretch}{1.32}
\caption{Performance for detecting bursts and inter-bursts with trac\'e alternant activity using 3 different classifiers.}
\centering
\resizebox{0.48\textwidth}{!}{
\begin{tabular}{rccc}
\toprule
& ~~FDA~~ & Random Forest & Linear-SVM \\
\midrule
\textbf{Median AUC} & 0.91 & 0.91 & 0.92 \\
\bottomrule
\end{tabular}
}
\begin{flushleft} \footnotesize{FDA: Fisher discriminant analysis; SVM: support vector machine} \end{flushleft}
\label{auc_vclass}
\end{table}

%The linear-SVM shows highest AUC performance and is selected for feature classification in proposed system pipeline. 
Testing performance of the inter-burst detector improves when a decision is made over the 4 channels. For a single channel, median AUC is 0.92 (95\% confidence intervals, CI: 0.90 to 0.97) compared to a median AUC of 0.95 (95\% CI: 0.93 to 0.98) from averaging the confidence score over all channels.

The post-processed envelope function is then tested to detect the presence of TA activity. Table \ref{taperf} presents the performance metrics for testing the TA detector using the held-out testing inter-burst detector model from the LOSO cross-validation. A threshold value of 2.06 on the envelope function gives approximately equal sensitivity and specificity (76.2\% and 76.3\%).

\begin{table}[!h]
\renewcommand{\arraystretch}{1.02}
\centering
\caption{Classification performance for the proposed TA detector.}
\label{taperf}
\resizebox{0.40\textwidth}{!}{
\begin{tabular}{cccc}
\toprule
~Cohen's~ & AUC & F1-score & ~Accuracy~ \\
kappa & (95\% CI)& (\%) & (\%) \\
\midrule
\multirow{2}{*}{0.57} & 0.84 & \multirow{2}{*}{79.7} & \multirow{2}{*}{81.9} \\
 & (0.80, 0.88) & & \\
\bottomrule
\end{tabular}
}
\begin{flushleft} \footnotesize{Key, AUC: Area-under receiver operating characteristic; CI: confidence interval.} \end{flushleft}
\vspace{-0.25cm}
\end{table}

% The temporal assessment between proposed TA detector and EEG expert annotation resulted in a kappa value of 0.58 and AUC value of 0.84 with sensitivity and specificity of 77.3\% and 78.0\%.

\section{Discussion and Conclusions}
\label{discussion}

We present a novel approach for detecting trac\'e alternant in the EEG of term neonates by first developing a burst--interburst detector and then post-processing this detector's confidence score. The method is trained and tested on a large EEG database from 72 term infants EEG, recorded within days of birth. The inter-burst detector, which combines amplitude and spectral characteristics using a machine learning approach, results in an AUC of 0.95 (95\% CI: 0.93 to 0.98). This detector is used to evaluate the presence of inter-burst intervals by generating an envelope which detects TA activity with reasonable performance ($\kappa = 0.57$, F1 = 79.7\%, accuracy = 81.9\%, and AUC = 0.84). 

%A previous study by Turnbull \emph{et al}. \cite{Turnbull} reported testing 20 TA segments for 6 full-term infants using different Wavelet transforms with additional post-processing. No detection results were presented.

It is difficult to directly compare our results to other studies. Most use EEG from both preterm and term infants \cite{Dereymaeker2,Ansari}, with some of the term EEG from infants born preterm \cite{Pillay,koolen}. The distinction in our work is that we developed our methods on EEG recorded within days of birth from a healthy cohort of term infants. Also, most methods focus on classifying different sleep states, such as active sleep and quiet sleep. Here we focus solely on TA activity, a subset of quiet sleep. Probably the most similar study to ours was presented by Turnbull \emph{et al.} \cite{Turnbull}, but the lack of detection results prevents direct comparison. 

%The method was developed and tested on 20 trac\'e alternant segments for 6 full-term infants. 
%This study used a general signal processing description of the burst and the inter-burst sequence but without a neurophysiological definition of these waveforms. 
 %Future work will include developing statistical feature set to evaluate repetitive nature of detected burst and the inter-burst pattern sequence to further improve classification performance.

In conclusion, we have developed a method to detect TA activity in the EEG of term neonates. By enabling simple spatio-temporal post-processing analysis on the burst--interburst sequence we generate promising system performance. Future work will investigate developing statistical features, in conjunction with machine learning methods, of the temporal organisation of the burst--interburst sequence to further improve classification performance. The proposed method could help highlight disturbances to the sleep--wake cycle by noting the presence or absence of TA activity and therefore help identify neurologically compromised infants, such as those with hypoxic-ischemic encephalopathies.

\end{document}